\pdfoutput=1
\documentclass[12pt]{article}
\usepackage{graphicx,nopageno,sectsty,hyperref}
\sectionfont{\normalsize}
\newcommand{\GeV}{\mbox{ GeV}}
\newcommand{\kt}{k_\perp}
\newcommand{\pt}{p_T}
\newcommand{\avktsq}{\langle k^2_\perp \rangle}
\newcommand{\avptsq}{\langle p^2_T \rangle}
\newcommand{\avgcos}{\langle \cos \varphi \rangle}
\newcommand{\zh}{z_h}
\newcommand{\xf}{x_F}
\newcommand{\Eh}{E_h}

\title{\normalsize \bf  AZIMUTHAL ASYMMETRY AND TRANSVERSE MOMENTUM IN DEEP
INELASTIC MUON SCATTERING\footnote{Proc. 7th Meeting of the APS DPF, Nov.~10--14, 1992, FNAL, Batavia, IL, Vol. 2, p932--934}}
\author{\small Mark D. Baker \\
\small \it Massachusetts Institute of Technology \\
\small \it for the Fermilab  E665 Collaboration \\
\small (2013 email address: mdbaker@bnl.gov) 
\date{}
}
\begin{document}
\maketitle
\begin{abstract}
The azimuthal asymmetry and the transverse momentum of forward produced charged
hadrons in deep inelastic muon scattering have been studied as a function of the 
event kinematics and of the hadron variables. Primordial $\kt$ of the struck 
parton and $\mathcal{O}(\alpha_s)$ corrections to the cross-section are expected
to contribute to the transverse momentum and the azimuthal asymmetry of hadrons. 
The data show some unexpected dependences not present in a Monte Carlo simulation 
which includes the theoretical parton-level azimuthal asymmetry.
\end{abstract}

\section{Introduction}

\vspace{-1.5 em}
\begin{figure}[h]
\centering 
\includegraphics[width=.8\textwidth]{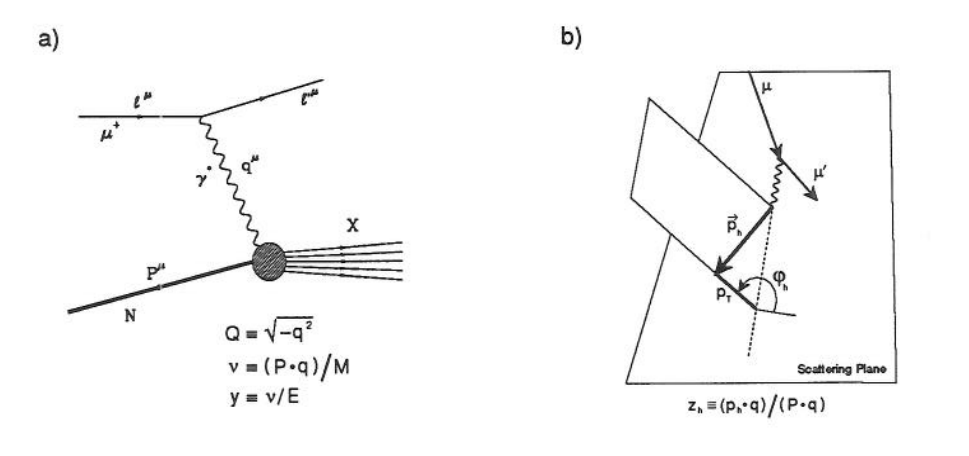}
\caption{\label{fig1}Deep Inelastic Muon Scattering.}
\end{figure}

Figure~\ref{fig1}a shows the leading order Feynman diagram for deep inelastic 
scattering and defines the kinematic variables. Figure~\ref{fig1}b shows the 
3-momentum vectors of the beam muon, scattered muon, and a typical hadron; 
the standard variables are also defined. In addition, we define the quantity 
\emph{Order} for each hadron such that the hadron with the highest $\zh$ is 
Order 1, the hadron with the second highest $\zh$ is Order 2 and so forth. 
The data presented in this talk were taken during the 1987 fixed target run 
at Fermilab by the E665 Collaboration using a 490~GeV muon beam and a deuterium 
target. The most important cuts were $Q^2 > 2 \GeV^2$, $\nu > 100 \GeV$, 
and $y < 0.9$.

Electromagnetic bremsstrahlung events, characterized by low neutral multiplicity,
and large neutral energy, were removed using an electromagnetic calorimeter. 
Charged hadrons were examined which had $\xf > 0.0$ and $\Eh > 6 \GeV$. Roughly 
$3 \times 10^4$ events and $6 \times 10^4$ hadrons survived these cuts. For 
further details on the event sample and the analysis, see 
Reference~\cite{Baker1993}.

In the Quark-Parton Model, an azimuthal asymmetry can arise due to primordial 
$\kt$ of the struck parton~\cite{Cahn1978,Cahn1989} or to $\mathcal{O}(\alpha_s)$ 
effects (gluon bremsstrahlung and photon-gluon fusion)~\cite{GP1978,JK1982}. 
If we assume that the primordial $\kt$ distribution is independent of $Q$, 
the theory predicts that the $\varphi$ asymmetry, dominated by the $\kt$, 
should disappear at large $Q$, going as $1/Q$. It is possible, however, that 
the effective value of $\kt$ depends upon $Q$ and $\nu$ (see, for instance, 
Reference~\cite{Land1977}). The theoretical parton-level $\varphi$ asymmetry 
was incorporated into a Monte Carlo~\cite{Baker1993} which was a modified version 
of Lepto 5.2 (Matrix Element) with $\avktsq = (0.44 \GeV)^2$ independent 
of $Q$. This Monte Carlo included a full simulation of the apparatus and used the 
same event reconstruction code that was used on data. Therefore, the Monte Carlo 
results should be compared to the uncorrected data.

\section{Results}

\begin{figure}[ht]
\centering 
\includegraphics[width=.9\textwidth]{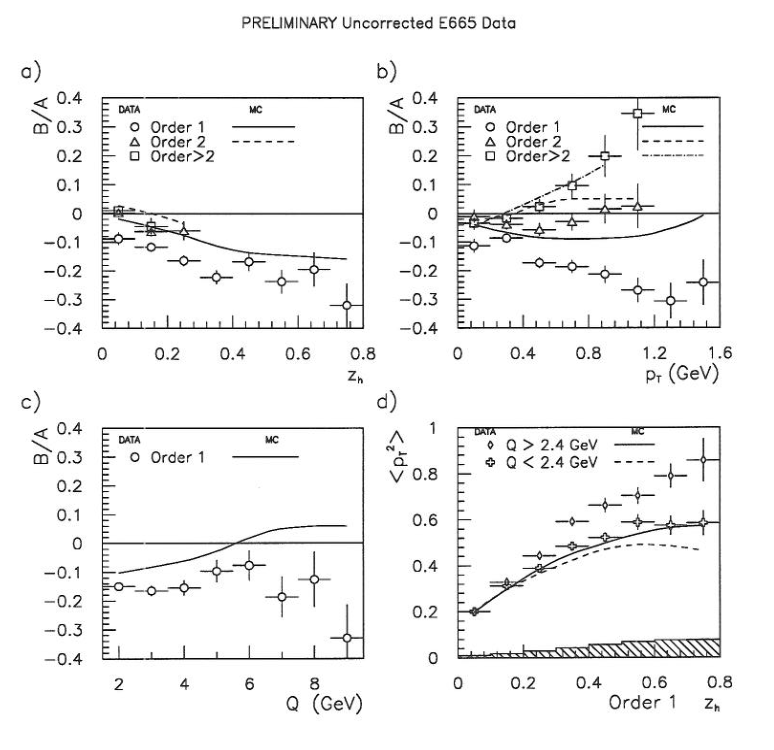}
\caption{\label{fig2} The data are shown overlaid with Monte Carlo curves. $B/A = 2 \avgcos$ is plotted 
for particles of various orders versus a) $\zh$, b) $\pt$, and c) $Q$; d) $\avptsq$ is plotted 
versus $\zh$ for Order 1 particles for high and low $Q$. In plot a), note that the Order 2 
particles are restricted kinematically to the range $\zh < 0.5$, and practically to the range 
$\zh < 0.3$. The data have not been corrected for acceptance; the Monte Carlo results include 
a simulation of the acceptance and should be compared to the uncorrected data.}
\end{figure}

In order to characterize the $\varphi$ asymmetry in the data, we fit $dN/d\varphi_h$
distributions to the form $A + B \cos \varphi_h$ and plot the quantity 
$B/A = 2 \avgcos$ as a function of $\zh$, $\pt$, or $Q$ for 
hadrons of various orders. The results are shown in Figure~\ref{fig2}a--c; the curves 
represent the Monte Carlo predictions. Since the effective $\kt$ may depend on $Q$ 
and since the $\kt$ contributes to hadron $\pt$ primarily at high $\zh$, it is 
interesting to compare the hadron $\avptsq$ distributions in the high $\zh$ 
region from events with “low $Q$” $(Q < 2.4 \GeV)$ to those with “high Q” 
$(Q > 2.4 \GeV)$. Figure 2d shows $\avptsq$ versus $\zh$ for Order 1 charged 
particles for both the low and the high $Q$ regions. The curves indicate the 
Monte Carlo predictions. The shaded region at the bottom of the plot indicates 
an estimate of the $\avptsq$ resolution and also the level of the systematic 
error on $\avptsq$.

The data show several effects that are not well described by the Monte Carlo, which uses 
the standard value $\avktsq = (0.44 \GeV)^2$, independent of $Q$. First, the $\varphi$ 
asymmetry in the data is mostly due to the Order 1 particle and depends on $\pt$ of the 
Order 1 hadron. Second, there is more $\varphi$ asymmetry and more $\pt$ in the data than 
in the Monte Carlo, possibly indicating that the overall $\avktsq$ is larger than 
$(0.44 \GeV)^2$. Finally, the $\varphi$ asymmetry in the data is independent of $Q$, 
persisting to higher values of $Q$ than expected, and there is more $\pt$ at high $Q$ 
than at low $Q$ in the data, especially for the high $\zh$ particles. These $Q$ dependences 
may be evidence that the effective $\avktsq$ depends on $Q$. It is also possible, of course, 
that our string model of hadronization or our treatment of the $\kt$ in the parton model 
is incorrect.

\end{document}